\documentclass[12pt]{article}
\usepackage{amsmath,amsfonts,amssymb,amsthm,amstext,eucal}
\usepackage{amscd,bbold}

\numberwithin{equation}{section}

\begin{document}
\newcommand{\todo}[1]{{\em \small {#1}}\marginpar{$\Longleftarrow$}}   
\newcommand{\labell}[1]{\label{#1}\qquad_{#1}} 
\newcommand{\ud}{\mathrm{d}}

\rightline{DCPT-05/51}   
\vskip 1cm

\begin{center} {\Large \bf Non-supersymmetric asymptotically AdS$_5
    \times S^5$ smooth geometries}
\end{center} 
\vskip 1cm   
  
\renewcommand{\thefootnote}{\fnsymbol{footnote}}   
\centerline{Simon 
F. Ross\footnote{S.F.Ross@durham.ac.uk}}    
\vskip .5cm   
\centerline{ \it Centre for Particle Theory, Department of  
Mathematical Sciences}   
\centerline{\it University of Durham, South Road, Durham DH1 3LE, U.K.}   
  
\setcounter{footnote}{0}   
\renewcommand{\thefootnote}{\arabic{footnote}}


\begin{abstract}
  We find soliton solutions in five-dimensional gauged supergravity,
  where a circle degenerates smoothly in the core of the
  geometry. In the family of solutions we consider, we find no
  completely smooth supersymmetric solutions, but we find discrete
  families of non-supersymmetric solitons. We discuss the relation to
  previous studies of the asymptotically flat case. We also consider
  gauged supergravities in four and seven dimensions, but fail to find
  any smooth solutions.
\end{abstract}

\section{Introduction}

There has recently been considerable interest in smooth supergravity
solutions, and their relation to the states of dual conformal field
theory (CFT) descriptions.  This was initiated in studies of the D1-D5
system~\cite{Balasubramanian:2000rt,Maldacena:2000dr}, where smooth
geometries whose near-horizon limit is global AdS$_3 \times S^3$ were
found. This work has been extended in a number of interesting ways, as
will be reviewed below. More recently, 1/2 BPS smooth asymptotically
AdS$_5 \times S^5$ solutions were found in~\cite{llm}.  Another class
of asymptotically AdS$_5$ solutions was recently found
in~\cite{Cvetic:2005zi,Chong:2005da,Chong:2005hr}. As
in~\cite{Balasubramanian:2000rt,Maldacena:2000dr}, these latter
solutions involve a degenerating circle, so they are more closely akin
to the D1-D5 system than to the bubbling AdS solutions of~\cite{llm}.
The aim of the present paper is to construct new examples of this type
in gauged supergravity, employing the approaches used in the study of
the D1-D5 system (the analysis will be closely based on the approach
of~\cite{nonbps}).

To begin with, let us briefly review the previous work in the D1-D5
system. In~\cite{Balasubramanian:2000rt,Maldacena:2000dr}, special
smooth solutions were found in the family of asymptotically flat
supersymmetric geometries~\cite{cy} describing a D1-D5 system wrapped
on a circle, with angular momentum in the transverse space. The
near-horizon limit of these metrics was global AdS$_3 \times S^3$.  By
studying the near-horizon limit, they were identified with Ramond-Ramond
ground states of the dual CFT.  They were generalised
in~\cite{lm1,lm2,lmm} to obtain a family depending on arbitrary
functions, corresponding to more general Ramond-Ramond ground states. These
were interpreted in~\cite{lmm} as Kaluza-Klein (KK) monopole
supertubes; the five-dimensional geometry has a KK monopole wrapped on
a contractible circle.

This construction was then generalised to include momentum along the
D1-D5 string
in~\cite{Lunin:2004uu,Giusto:2004id,Giusto:2004ip,nonbps,Bena:2005va,Berglund:2005vb}.
The resulting five-dimensional geometries are smooth up to some
orbifold singularities, and there are again large families of
solitons. The CFT interpretation of the most general solutions in this
class has not yet been understood.  In~\cite{nonbps}, the first
non-supersymmetric solitons were found, by returning to the general
family of metrics in~\cite{cy} and considering the restrictions on the
parameters necessary to ensure smoothness.  States of the dual CFT
corresponding to the AdS$_3 \times S^3$ geometries obtained in the
near-horizon limit were also identified. It is really remarkable that
such non-supersymmetric solitons exist, and further exploration of
them may provide important insights into the relation between CFT
states and geometry.

In this paper, we will find new non-supersymmetric soliton solutions
by applying the approach used in~\cite{nonbps} to gauged supergravity.
That is, we consider the families of five-dimensional asymptotically
AdS$_5$ charged rotating black holes found
in~\cite{clp1,clp2,Chong:2005da,Chong:2005hr}, and ask what
restrictions on the parameters are implied by requiring that the
geometry closes off smoothly in the interior with a degenerating
circle. This problem was already considered for the supersymmetric
limit of the solutions
in~\cite{Cvetic:2005zi,Chong:2005da,Chong:2005hr}, where special cases
that give smooth metrics were identified. We generalise this by
considering the general non-supersymmetric solutions
of~\cite{clp1,clp2,Chong:2005da,Chong:2005hr}, and we consider the
conditions necessary to ensure the matter fields are also smooth.  The
latter conditions turn out not to be satisfied by any of the
supersymmetric solutions
in~\cite{Cvetic:2005zi,Chong:2005da,Chong:2005hr}.\footnote{There is a
  component of the gauge potential along the degenerating direction
  which does not vanish where the circle degenerates.  This
  corresponds to twisting the degenerating circle over a circle in the
  $S^5$ by a non-integer amount, in the description of this gauged
  supergravity as arising from Kaluza-Klein reduction of IIB on an
  $S^5$.}  That is, once we impose these gauge field smoothness
conditions, we have no smooth supersymmetric solitons. There are
discrete families of smooth non-supersymmetric solitons which
satisfy all these conditions.\footnote{There are other asymptotically
  locally AdS solitons which have been constructed by different
  methods: The AdS soliton of~\cite{Horowitz:1998ha} asymptotically
  approaches a $\mathbb{Z}$ quotient of AdS in Poincare coordinates,
  and other asymptotically AdS$/\mathbb{Z}_p$ solitons were
  recently constructed in~\cite{eh1,eh2}.}

The bulk of the paper is taken up with the analysis of the different
families of solutions from~\cite{clp1,clp2,Chong:2005da,Chong:2005hr}.
In the next section, we start by considering the simplest case, a
charged rotating black hole in minimal gauged supergravity with three
equal charges and two equal angular momenta. We then extend this to
unequal charges, in section~\ref{uneqch}, or to unequal angular
momenta, in section~\ref{uneqang}. (Solutions with both unequal
charges and unequal angular momenta are not yet available in the
literature.) In section~\ref{poinc}, we obtain solutions which are
asymptotically AdS$_5$ in Poincar\'e coordinates by taking a large
mass limit of the previous solutions. The same geometries can be
obtained by double analytic continuation of charged black holes. In
section~\ref{gto0}, we consider the limit of vanishing cosmological
constant, and discuss the relation to the asymptotically flat
solutions of~\cite{nonbps}.

The extension from asymptotically flat to asymptotically AdS$_5$
boundary conditions is interesting for two main reasons. It offers the
possibility of finding some dual description of these solutions in a
four-dimensional CFT, living on the boundary of AdS$_5$. (However, the
fact that all the solitons we find are non-supersymmetric makes it
more difficult to obtain a CFT description.) It is also interesting
because black holes in AdS are better behaved than black holes in
asymptotically flat space. If we think of these geometries as
describing microstates of black holes, following the proposal of
Mathur and collaborators~\cite{lm1,lm2} (see also the
review~\cite{mrev}), then the asymptotically AdS$_5$ case may make it
easier to make quantitative comparisons.
 In section~\ref{concl}, we offer some concluding comments on these
 and other issues and briefly discuss the extension of the approach
 used here to gauged supergravities in other dimensions. We consider
 the four and seven dimensional theories, and argue that the known
 families of charged rotating black hole solutions do not contain any
 solitons as special cases. 

\section{Equal charges and equal angular momenta}
\label{equal}

We start from the solution of~\cite{clp1}, describing a charged
rotating black hole in minimal five-dimensional gauged supergravity.
We work in the parametrisation introduced in~\cite{uniq}, but write
the cosmological constant as $\lambda = -g^2$. The metric is
\begin{eqnarray} \label{eqsol}
\mathrm{d}s^2&=&-\frac{r^2
W}{4b^2}\mathrm{d}t^2+\frac{1}{W}\mathrm{d}r^2 +\frac{r^2}{4}
(\sigma_1^2+\sigma_2^2)+b^2(\sigma_{3}+f\mathrm{d}t)^2,
\\ A&=&\frac{\sqrt{3}q}{r^2}\left[\mathrm{d}t-\frac{1}{2}j \sigma_{3}\right],
\end{eqnarray}
where
\begin{eqnarray} 
  b^2&=&\frac{1}{4r^4}\left[r^6-j^2
    q^2+2j^2 p r^2\right], \\
  f&=&-\frac{j}{2b^2}\left[\frac{2p-q}{r^2}-\frac{q^2}{r^4}\right],\\
  W&=&1+ g^2 r^2+\frac{1}{r^2}\left[2j^2 g^2p- 2(p-q)\right]
  +\frac{1}{r^4}\left[q^2+j^2( 2p-g^2 q^2)\right] , 
\end{eqnarray} 
and the $\sigma_i$ are the left-invariant one-forms on $S^3$,
\begin{eqnarray}
\sigma_1 &=& \cos \psi d \theta + \sin \psi \sin \theta d\phi,
\label{sigma1} \\
\sigma_2 &=& -\sin \psi d \theta + \cos \psi \sin \theta d\phi, \\
\sigma_3 &=& d\psi + \cos \theta d\phi. \label{sigma3}
\end{eqnarray}

The determinant of the metric is $g = -r^6/16$, so the only potential
singularities are at $W=0$, $b=0$ (where $f$ diverges) and $r=0$.
Since the determinant goes like $r^6$, the singularity at $r=0$ is a
real singularity, unlike in the analysis of the D1-D5 case
in~\cite{nonbps}, where it could be removed by a coordinate
transformation. The determinant of the metric on a surface of constant
$r$ is $g_{|r=r_0} = -r_0^6 W/16$, so the singularity at $W=0$
corresponds either to a horizon or to a degeneration in the spatial
metric. The determinant of the spatial metric is $g_{|r=r_0, t=t_0} =
r^4 b^2/4$, so there is a degeneration in the spatial metric if $b=0$
as well.

Thus, to have a smooth solution with a circle degenerating in the
interior, there needs to be some $r=r_0>0$ which is, firstly, the
largest root of $W$, so $W(r_0)=0$ and $W(r) >0$ for $r > r_0$.
Secondly, it must be the largest root of $b$, so $b^2(r_0)=0$ and
$b^2(r)>0$ for $r> r_0$.  Thirdly, there must be no $dt
d\psi$ cross terms at $r=r_0$, as they would prevent us from
constructing a smooth Cartesian coordinate system near this origin;
this requires $b^2 f(r_0) = 0$, or equivalently that $f$ be finite at $r=r_0$. 
This final condition is the easiest one to solve, and gives us
\begin{equation} \label{r0def}
r_0^2 = \frac{q^2}{2p-q}. 
\end{equation}
Requiring that $W(r_0)=0$ and $b^2(r_0)=0$ then imposes conditions on
the parameters; these conditions are both satisfied if
\begin{equation} \label{jcond}
j^2 = - \frac{q^3}{(2p-q)^2},
\end{equation}
which requires $q<0$.\footnote{Note that the situation is not
  symmetric under $q \to -q$ because of the presence of a Chern-Simons
  term in the gauged supergravity Lagrangian.} The BPS bound is then
$p \geq 0$. This $r_0$ is automatically the largest root of $b^2$.
Requiring it to be the largest root of $W$ implies
\begin{equation} \label{w}
w \equiv g^2 q^3 + 2p(2p-q) <0.  
\end{equation}

With this choice of parameters, as $r \to r_0$, the proper size of the
$\psi$ circle goes to zero. An important difference relative to the
case studied in~\cite{nonbps} is that the period of $\psi$ is already
fixed; to have an asymptotically AdS$_5$ spacetime, $\Delta \psi =
4\pi$. More generally, $\phi,\psi,\theta$ could define a smooth
quotient of $S^3$ asymptotically, which would permit $\Delta \psi =
4\pi/k$ for integer $k$. (Our main interest will be in the
asymptotically AdS$_5$ case $k=1$.) Because the periodicity of $\psi$
is already fixed, the condition that $r=r_0$ is a smooth origin in the
plane is going to impose an additional restriction on the parameters.
Near $r=r_0$, the relevant part of the metric is
\begin{equation}
ds^2 = \ldots + \frac{(2p-q)q^2 dx^2}{(3q-2p) w} + \frac{3q-2p}{4q} x^2
d\psi^2, 
\end{equation}
where $x^2 \equiv r^2 -r_0^2$, and $w$ is given in \eqref{w}.  We must
then have $3q-2p <0$ and $w <0$ for the metric to have the correct
signature.  The regularity condition that $r=r_0$ be a smooth origin
then requires
\begin{equation} \label{regcond}
\frac{(2p-3q)^2}{(2p-q)q^3} \left( g^2 q^3 +
  2p(2p-q)\right) = k^2. 
\end{equation}

When the two conditions \eqref{jcond} and \eqref{regcond} are
satisfied, the metric is completely smooth, with a smooth origin at $r
= r_0$. Since $W > 0$ for $r > r_0$, the surfaces of constant $r$ are
timelike; hence, there are timelike curves that move to $r \to \infty$
from any point in the spacetime, showing that there are no event
horizons in this spacetime. Since $g^{tt} = -
4b^2/(r^2 W) < 0$ for $r \geq r_0$, $t$ is a global time function for
our soliton, and the metric is stably causal (and hence free of
closed timelike curves).

So far we have imposed two conditions on the parameters, \eqref{jcond} and
\eqref{regcond}. This leaves one free parameter in this class of
solutions. However, there is another regularity condition that we need
to impose: regularity of the matter fields. The gauge potential near
$r=r_0$ is  
\begin{equation}
A=\frac{\sqrt{3}(2p-q)}{q}\mathrm{d}t-\frac{\sqrt{-3q}}{2}
\sigma_{3},
\end{equation}
so $A_\psi$ does not vanish near $r=r_0$. The holonomy around this
direction, 
\begin{equation} 
\oint_{S^1} A = -\frac{\sqrt{-3q}}{2} \Delta \psi, 
\end{equation}
is a gauge invariant quantity. This holonomy seems to provide an
obstruction to making $r=r_0$ a regular origin; If $r=r_0$ was a
regular origin this circle would be a contractible cycle, and one
would expect
\begin{equation}
\oint_{S^1} A = \int_{D^2} F \to 0
\end{equation}
as the circle contracts. However, the gauge potential $A$ comes from
a diagonal $U(1) \subset U(1)^3 \subset SO(6)$. Since the gauge group
is compact, the holonomy takes values in a circle, and can be shifted
by an integer. Thus, regularity only requires $\oint_{S^1} A
\to 0$ modulo an appropriate period.

The issue is perhaps most easily understood by recalling that from the
ten-dimensional perspective, these are Kaluza-Klein gauge fields, and
the gauge field $A$ comes from writing a Kaluza-Klein
ansatz~\cite{Cvetic:1999xp}
\begin{equation}
ds^2_{10} = ds^2_{5} + \frac{1}{g^2} \sum_i [d\mu_i^2 + \mu_i^2 (d\phi_i +
\frac{g}{\sqrt{3}} A)^2 ],
\end{equation}
where $\mu_i, \phi_i$ $i = 1,\ldots,3$ are coordinates on an $S^5$, so
$\sum_i \mu_i^2=1$ and $\Delta \phi_i = 2\pi$. There are globally
well-defined large coordinate transformations in this metric, $\phi \to
\tilde \phi = \phi + \frac{km}{2} \psi$ for integer $m$. These coordinate
transformations preserve the periodicity of the coordinates; that is,
$\tilde \phi \sim \tilde \phi + 2 \pi$ and $\psi \sim \psi + 4\pi/k$
generate the same identifications as $\phi \sim \phi + 2 \pi$ and
$\psi \sim \psi + 4\pi/k$. From the
Kaluza-Klein perspective, these coordinate transformations correspond
to large gauge transformations $A \to A +
\frac{\sqrt{3}km}{2g} d\psi$. They can therefore be used to remove the
non-zero $A_\psi$ near $r=r_0$ if
\begin{equation} \label{qquant}
\sqrt{-q} = \frac{km}{g} .
\end{equation}
That is, the condition on the holonomy is $\oint_{S^1} A \to 0$ mod
$2\pi \sqrt{3}/g$, which is satisfied if \eqref{qquant} is. This
condition tells us that the twisting of the degenerating circle on the
$S^5$ is quantized. Thus $m$ is analogous to the integers specifying
the smooth solitons in~\cite{nonbps}.

Thus, for a given asymptotic boundary condition, there is a
one-integer parameter family of smooth solitons. In particular, for
the asymptotically AdS$_5 \times S^5$ solutions, $k=1$, and the
solitons are labelled by $m =1,2,\ldots$. In table~\ref{t1}, the
values of the parameters for the first few solitons are listed. Note
that there is no supersymmetric soliton; that is, there is no case
with $p=0$.\footnote{In particular, the equal-charge case of the
  supersymmetric solutions in~\cite{Cvetic:2005zi} does not
  satisfy~\eqref{qquant}.  This was noted indirectly
  in~\cite{Cvetic:2005zi}, where it was observed that the correct
  periodicity conditions for the gravitini could not be satisfied for
  the solutions considered there. The relation to~\cite{Cvetic:2005zi}
  will be discussed in more detail in the next section.}

\begin{table}
\begin{center}
\begin{tabular}{|c||c|c|c||c|} \hline
$m$ & $p$ & $q$ & $j$ & $M$  \\ \hline \hline
1 & 0.281 &-1 & 0.640 & 3.108 \\
2 & 3.063 &-4 & 0.790 & 18.143 \\
3 & 11.351  &-9 &0.852 & 54.416 \\
4 & 28.143 &-16 & 0.885 & 121.337  \\ \hline
\end{tabular}
\end{center}

\caption{The values of the parameters and thermodynamic mass for the
  first few solitons, for $k=1$. All quantities are quoted in units in which
  $g=1$.}
\label{t1}
\end{table}

Since our parametrisation is not particularly physical, it is more
useful to give the asymptotic charges. It is clear from the form of
the gauge potential that the conserved charge is $q$, up to a
normalisation factor. The angular momentum $J$ can be calculated using
the Komar integral technique and is given by
\begin{equation} \label{angmom}
J =\frac{\pi}{4}j (2p-q) = \frac{\pi}{4 g^3} k^3 m^3,
\end{equation}
so the angular momentum has a simple expression in terms of $m$. A
thermodynamic mass was obtained for this general family of black holes
in~\cite{uniq}:
\begin{equation}
M =\frac{\pi}{4}\left( 3(p-q)+p j^2 g^2 \right).
\end{equation}
This does not have such a simple form in terms of the basic
parameters, but its value for the first few solitons is listed in
table 1. 

An interesting difference between these solitons and the
asymptotically flat ones discussed in~\cite{nonbps} is that, as
suggested in~\cite{nonbps}, they do not have an ergoregion. That is,
the Killing vector $V = \partial_t$ is everywhere timelike or null. Writing
$r^2 = r_0^2 + z^2$, its norm is 
\begin{equation}
g_{\mu\nu} V^\mu V^\nu = - \frac{z^2}{4b^2 (r_0^2+z^2)^4} P_{10}(z), 
\end{equation}
where $P_{10}(z)$ is a tenth-order even polynomial in $z$. The
coefficients in this polynomial are all positive when $r_0^2$ is given
by~\eqref{r0def} and $j^2$ is given by~\eqref{jcond}, so the norm is
non-positive for $z^2 \geq 0$, that is, for $r^2 \geq r_0^2$. The
absence of an ergoregion suggests that these geometries stand a better
chance of being stable than the solitons in~\cite{nonbps}. It would
clearly be interesting to explore this issue in more detail.

To close this section, we make a remark about spin structures on these
solutions for the case $k=1$.\footnote{This issue is discussed in
  detail in the revised version of~\cite{Cvetic:2005zi}.} The surfaces
of constant $r$ have topology $\mathbb{R} \times S^3$, so they have a
unique spin structure. The metric on $S^3$ can be written as
\begin{eqnarray}
ds^2_{S^3}  &=& \frac{r^2}{4} (d\theta^2 + \sin^2 \theta d\phi^2) +
\frac{b^2}{4} (d\psi + \cos \theta d\phi)^2 \nonumber \\ 
&=&  r^2 (d\vartheta^2 + \sin^2 \vartheta \cos^2 \vartheta
(d\phi_1 - d\phi_2)^2) + b^2(\cos^2 \vartheta d\phi_1 + \sin^2
\vartheta d\phi_2)^2,
\end{eqnarray}
where $\theta = 2\vartheta$, $\psi = \phi_1 + \phi_2$, $\phi = \phi_1
- \phi_2$. Now the $\phi_{1,2}$ circles are contractible (they shrink
to zero at $\theta = 0,\pi$ respectively), so the spin structure must
have antiperiodic fermions around these circles. This implies the
fermions are periodic around the $\psi$ circle. But this does not
define a consistent spin structure on the full spacetime, as the
$\psi$ circle is also contractible: it degenerates at $r=r_0$.  Thus,
the five-dimensional solutions are not spin manifolds.

However, this problem is easily resolved in the ten-dimensional
geometry. The $\psi$ circle at fixed position on the $S^5$ has a
periodic spin structure by the above argument. But the circle which
degenerates smoothly in the interior is the $\psi$ circle at fixed
values of $\tilde \phi_i = \phi_i + \frac{m}{2} \psi$; that is, it
corresponds to going around the $\psi$ circle while going around each
of the $\phi_i$ circles $m$ times. Since the fermions must also be
antiperiodic around the $\phi_i$ circles, they will be antiperiodic
around the degenerating cycle, as required, for odd values of $m$.
Thus, for odd $m$, the ten-dimensional spacetime has a unique
well-defined spin structure. From the five-dimensional point of view,
the presence of the Kaluza-Klein gauge field allows us to define a
spin$^c$ structure on the spacetime even though it does not have a
spin structure. That is, the fermions taking values in this spin
bundle on the ten dimensional spacetime Kaluza-Klein reduce to give us
spinor fields in five dimensions charged under the gauge
field.

\section{Unequal charges and equal angular momenta}
\label{uneqch}

The equal-charge solution studied above was generalised to unequal
charges in~\cite{clp2}. This more general family also contains smooth
solitons, but there are still no supersymmetric cases,
contrary to the claims in~\cite{Cvetic:2005zi}.  The generalisation
means that the solitons will be labelled by three integer parameters
determining the charges, replacing the single parameter $m$ in the
above solutions.

We start from the solutions of~\cite{clp2} in the form introduced
in~\cite{uniq}, where the metric and gauge fields are 
\begin{equation} 
\mathrm{d}s^2= -\frac{RY}{f_{1}}\mathrm{d}t^2 +
\frac{R\bar \rho^2}{Y}\mathrm{d}\bar \rho^2 +
\frac{1}{4}R(\sigma_1^2+\sigma_2^2) + \frac{f_{1}}{4R^2}
(\sigma_3-2\frac{f_2}{f_1}\mathrm{d}t)^2,
\end{equation}
and
\begin{equation} 
A^i = \frac{r_j r_k - \Gamma r_i}{ \bar \rho^2 \tilde H_i} dt  +
\frac{L r_i}{2 \bar \rho^2 \tilde{H}_i}  \sigma_3,
\end{equation}
with
\begin{eqnarray}
f_1&=&R^3+L^2\left( \bar \rho^2-\frac{1}{3}\sum_{i}r_{i}^2 \right), \\
f_2&=&L \left( \Gamma \bar \rho^2-\frac{1}{3}\Gamma\sum_{i}r_{i}^2+
r_1 r_2 r_3 \right), \\
Y&=&-\lambda R^3+\bar \rho^4+\Big(\frac{1}{3}\sum_{i}r_{i}^2-\lambda L^2
-\Gamma^2 \Big)\bar \rho^2+\frac{1}{3}\Gamma^2\sum_{i}r_{i}^2-2\Gamma
  r_1 r_2 r_3\nonumber \\&& +\frac{1}{3}\lambda L^2
  \sum_{i}r_{i}^2+L^2+\left[\frac{5}{18}(\sum_{i}r_{i}^2)^2-\frac{1}{2}\sum_{i}r_{i}^4\right],  
\end{eqnarray}
and
\begin{equation}
R = \bar \rho^2 \Big(\prod_{i}^3
\tilde{H}_{i}\Big)^\frac{1}{3}, \quad
\tilde{H}_{i}=1+\frac{2r_i^2 - r_j^2 - r_k^2}{3\bar \rho^2}.
\end{equation}
Because the charges are unequal, this solution also involves
non-trivial scalar fields $\varphi_1, \varphi_2$, given by
\begin{equation}
e^{\frac{2}{\sqrt{6}} \varphi_1} = X_3, \quad e^{\sqrt{2} \varphi_2} =
  \frac{X_2}{X_1}, \quad X_i = \frac{R}{\bar \rho^2 \tilde{H}_i}. 
\end{equation}

The parametrisation of the solutions above was introduced
in~\cite{uniq} to demonstrate the uniqueness of the solutions.
However, it is not the most physically transparent or convenient
parametrisation, so before searching for smooth solitons, we would
like to introduce a more convenient parametrisation and a minor
coordinate transformation. Introduce
\begin{equation}
\ell_1^2 = \frac{r_1 r_2 r_3}{\Gamma}, \quad \ell_2^6 = L^2 \ell_1^2,
\quad q_i =
\frac{r_i^2}{\ell_1^2} - 1,
\end{equation}
set as usual $\lambda = -g^2$, and shift the radial coordinate,
\begin{equation}
\bar \rho^2 = \rho^2 + \frac{\ell_1^2}{3}(q_1 + q_2 + q_3).
\end{equation}
The powers are chosen so that $\ell_2$ and $\ell_1$ have length
dimensions, and $q_i$ are dimensionless. In this parametrisation,
\begin{equation} \label{uncmet}
\mathrm{d}s^2= -\frac{RY}{f_{1}}\mathrm{d}t^2 +
\frac{R\rho^2}{Y}\mathrm{d}\rho^2 +
\frac{1}{4}R(\sigma_1^2+\sigma_2^2) + \frac{f_{1}}{4R^2}
(\sigma_3-2\frac{f_2}{f_1}\mathrm{d}t)^2,
\end{equation}
with
\begin{eqnarray}
f_1 &=& R^3 + \frac{\ell_2^6}{\ell_1^2} \rho^2 - \ell_2^6,\\
f_2 &=& \ell_2^3 \sqrt{q_1+1} \sqrt{q_2+1} \sqrt{q_3+1} \; \rho^2, \\
Y - g^2 f_1 &=& \rho^4 - \rho^2 \ell_1^2 \left[
  q_1 q_2 + q_1 q_3 + q_2 q_3 + q_1 q_2 q_3 \right]+
\frac{\ell_2^6}{\ell_1^2} - \ell_1^4 q_1 q_2 q_3, 
\end{eqnarray}
and $R^3 = \bar{H}_1 \bar{H}_2 \bar{H}_3$, where
\begin{equation}
\bar{H}_i = \bar \rho^2 \tilde{H}_i = \rho^2 + \ell_1^2 q_i,
\end{equation}
and the gauge field becomes
\begin{equation} \label{uncgauge}
  A^i = - q_i \sqrt{q_j+1} \sqrt{q_k+1} \frac{\ell_1^2}{(\rho^2
    +\ell_1^2 q_i)} dt + \sqrt{q_i+1} \frac{\ell_2^3}{2(\rho^2 +
    \ell_1^2 q_i)} \sigma_3. 
\end{equation}
Thus, the parameters $q_i$ are proportional to the gauge charges. 

Let us now investigate the conditions for a smooth origin. As in the
previous case, there will be a smooth origin where the $\psi$ circle
degenerates at $\rho^2 = \rho_0^2$ if both the determinant of the
metric on surfaces of constant $r$ vanishes there and the determinant
of the metric on surfaces of constant $r$ and $t$ vanishes there. This
requires that $\rho^2 = \rho_0^2$ is the largest root of $f_1$ and
$Y$. This must also be a root of $f_2$ to eliminate cross terms, and
we must have $\bar{H}_i (\rho_0) > 0$ to ensure there are
no singularities elsewhere. Requiring that $\rho_0$ is a root of $f_2$
determines $\rho_0^2=0$, so $\bar{H}_i (\rho_0) > 0$ implies $\ell_1^2
q_i >0$. As in the equal charge case, this corresponds to negative
gauge charges. Requiring $f_1(\rho =0) = 0$ implies
\begin{equation} \label{bcond}
 \ell_2^6 = R^3(\rho=0) = \ell_1^6 q_1 q_2 q_3,
\end{equation}
and it is easy to see that  $Y(\rho=0)= 0$ as
well. This is automatically the largest root of $f_1$. It will be the
largest root of $Y$ as well if $g^2 \ell_1^2 > 1$. 

We now have the proper size of the $\psi$ circle going to zero at
$\rho = 0$. As in the previous section, the period of the $\psi$
circle is already fixed by imposing regularity of the metric on the
spheres of constant $\rho$ and $t$ at $\rho > 0$. To have a
smooth origin, we therefore need to impose an additional condition on
the parameters. Near $\rho= 0$, the relevant part of the metric
is
\begin{equation}
ds^2  = \ldots + \frac{R}{\ell_1^2} \left[ \frac{d\rho^2}{(q_1
    q_2+q_1 q_3+ q_2 q_3 + q_1q_2q_3)(g^2 \ell_1^2 - 1)} +\frac{ (q_1
    q_2+q_1 q_3+ q_2 q_3 + q_1q_2q_3) }{4q_1 q_2 q_3} \rho^2
  d\psi^2\right].
\end{equation}
Thus, since $\Delta \psi = 4\pi/k$, the regularity condition is
\begin{equation} \label{percond}
\frac{ (q_1 q_2+q_1 q_3+ q_2 q_3 +
  q_1q_2q_3)^2(g^2 \ell_1^2-1)}{q_1 q_2 q_3} = k^2. 
\end{equation}

When \eqref{bcond} and \eqref{percond} are satisfied, the metric is
smooth. As in the equal charge case, the surfaces of constant $\rho$ are
timelike, so there is no event horizon, and $g^{tt} <0$, so $t$ is a
global time function, and there are no closed timelike curves. We have
not checked whether any of these geometries have ergoregions, but we do
not expect that they would. 

Finally, we have to require that we can set $A_\psi^i =0$ at $\rho=
0$ by an allowed gauge transformation. In the gauge used above,
\begin{equation}
A_\psi^i(\rho=0) = \frac{\ell_1}{2} \left( \frac{q_j q_k (1+q_i)}{q_i}
\right)^{1/2}. 
\end{equation}
These gauge fields come from cross terms $(d\phi_i + g A^i)^2$ in a
Kaluza-Klein reduction from ten dimensions~\cite{Cvetic:1999xp} (note
there is a small difference in normalisation compared to the previous
section), so the allowed large gauge transformations are $A^i \to A^i
+ km^i d\psi/2g $.  The regularity condition is therefore
\begin{equation} \label{mi}
\frac{q_j q_k (1+q_i)}{q_i} =  \frac{m_i^2 k^2}{g^2 \ell_1^2}
\end{equation}
where $m_i$ are integers. The family of solitons for given asymptotic
conditions are then specified by the three integers $m_i$. We have not
attempted to solve (\ref{bcond},\ref{percond},\ref{mi}) explicitly to
determine the parameters in terms of these integers.

The discussion will simplify if we consider supersymmetric solutions.
Translating from the parametrisation in \cite{Cvetic:2005zi} to our
parametrisation, the BPS limit referred to as case A in
\cite{Cvetic:2005zi} is reached by taking $\ell_1 \to \infty$. To have
finite gauge charges, $q_i \ell_1^2$ must remain finite in the limit;
let us define $\bar q_i = q_i \ell_1^2$. $\ell_2$ is also fixed in
this limit by \eqref{bcond}. Then the regularity conditions for the
BPS case reduce to 
\begin{equation} \label{bps1}
\ell_2^6 = \bar q_1 \bar q_2 \bar q_3,
\end{equation}
\begin{equation} \label{bps2}
\frac{ g^2 (\bar q_1 \bar q_2+ \bar q_1 \bar q_3+ \bar q_2 \bar
  q_3)^2}{ \bar q_1 \bar q_2 \bar q_3} = k^2, 
\end{equation}
and
\begin{equation} \label{bps3}
\frac{\bar q_j \bar q_k}{\bar q_i} = \frac{m_i^2 k^2}{g^2}.
\end{equation}
The first condition \eqref{bps1} corresponds to the `critical
rotation' condition $\alpha^2 = q_1 q_2 q_3$ in~\cite{Cvetic:2005zi},
and \eqref{bps2} corresponds to equation (3.33)
in~\cite{Cvetic:2005zi}. The additional condition arising from
imposing regularity of the gauge field~\eqref{bps3} was not considered
previously.

We can rewrite \eqref{bps2} as
\begin{equation}
g^2 \left( \sqrt{\frac{\bar q_1 \bar q_2}{\bar q_3}} +
  \sqrt{\frac{\bar q_2 \bar q_3}{\bar q_1}} + \sqrt{\frac{\bar q_1
      \bar q_3}{\bar q_2}} \right)^2 = k^2. 
\end{equation}
Substituting \eqref{bps3} into this equation, we
find
\begin{equation}
|m_1| + |m_2| + |m_3| = 1,
\end{equation}
which cannot be satisfied, as the $m_i$ are all nonzero integers. Note
that setting some $m_i =0$ would correspond to setting some of the
gauge charges to zero, which will not produce a regular solution, as
it would imply $\bar{H}_j(\rho=0)=0$, and one of the scalar fields
would blow up at the origin. Thus, as in the equal-charge case, there
are no supersymmetric cases which satisfy all the regularity conditions.

\section{Equal charges and unequal angular momenta}
\label{uneqang}

All the solutions we have discussed so far have the two angular
momenta equal. Recently, a solution with unequal angular momentum
parameters was constructed~\cite{Chong:2005da,Chong:2005hr} (with
equal charges; the most general case, unequal angular momenta and
unequal charges, has not yet been found). Are there more smooth
solitons as special cases of this solution? It would seem unlikely
that this is possible, since for a general choice of angular momenta,
there is no $S^1$ in the $S^3$ whose size is independent of $\theta$
at all $r$.  Thus, it would seem difficult to choose parameters so
that we get a smooth origin for all $\theta$.  Remarkably, this can be
achieved, and there are smooth solitons.

For this case, the metric and gauge fields are  
\begin{eqnarray} \label{uneqmet}
ds^2 &=& - \frac{\Delta_\theta [ (1+g^2 r^2) \rho^2 dt + 2q\nu]
  dt}{\Xi_a \Xi_b \rho^2} + \frac{2q\nu \omega}{\rho^2} +
\frac{f}{\rho^4} \left( \frac{\Delta_\theta dt}{\Xi_a \Xi_b} - \omega
\right)^2 \nonumber \\ &&+ \frac{\rho^2 r^2 dr^2}{\tilde \Delta_r} +
\frac{\rho^2 d\theta^2}{\Delta_\theta}  + \frac{r^2+a^2}{\Xi_a} \sin^2 \theta
d\phi^2 + \frac{r^2+b^2}{\Xi_b} \cos^2 \theta d\psi^2,
\end{eqnarray}
and
\begin{equation}
A = \frac{\sqrt{3}q}{\rho^2} \left( \frac{\Delta_\theta dt}{\Xi_a
    \Xi_b} - \omega \right),
\end{equation}
where
\begin{eqnarray}
\nu &=& b \sin^2 \theta d\phi + a \cos^2 \theta d\psi, \\
\omega &=& a \sin^2 \theta \frac{d\phi}{\Xi_a} + b \cos^2 \theta
\frac{d\psi}{\Xi_b}, \\
\Delta_\theta &=& 1 - a^2 g^2 \cos^2 \theta -
b^2 g^2 \sin^2 \theta, \\
\tilde \Delta_r &=& (r^2+a^2)(r^2+b^2)(1+g^2 r^2) + q^2 + 2abq
- 2M r^2, \\
f &=& \left(2M + 2 abq g^2 \right) \rho^2 -q^2, \\
\rho^2 &=& r^2 + a^2 \cos^2 \theta + b^2 \sin^2 \theta, \label{rhosq} \\
\Xi_a &=& 1 - a^2 g^2, \quad \Xi_b  = 1 - b^2 g^2. 
\end{eqnarray}
The determinant of the metric is
\begin{equation}
\det g = - \frac{ \rho^4 r^2 \sin^2 \theta \cos^2 \theta}{\Xi_a^2
  \Xi_b^2}.  
\end{equation}
If $a=b$, this reduces to the solution in section~\ref{equal}, after
performing the coordinate transformation and redefinition of
parameters
\begin{equation}
\bar{r}^2 = \frac{r^2 +a^2}{\Xi_a}, \quad \bar j = a, \quad \bar q =
\frac{q}{\Xi_a^2}, \quad p = \frac{M+q}{\Xi_a^3}. 
\end{equation}

We want to find the conditions under which the solution has a smooth
origin at some value of $r$, where the orbits of a Killing vector
\begin{equation}
\xi = m \partial_\phi + n \partial_\psi
\end{equation}
go smoothly to zero size. We will state the conditions without giving
the sometimes tedious details of their derivation. We can only satisfy
the conditions at a negative value of $r^2$; we anticipate this by
writing the origin as $r^2 = -r_0^2$. We must take $m$ and $n$ to be
integers, so that these orbits are closed circles, and assume $mn \neq
0$, so the orbits do not have zero size at some point on the $S^3$ at
generic $r$. In this section, we will only consider the asymptotically
AdS case, corresponding to $k=1$ in the previous sections. We
therefore take $m$ and $n$ to be relatively prime. We can then define
another Killing vector
\begin{equation}
\chi = k \partial_\phi + l \partial_\psi
\end{equation}
with $k$ and $l$ integers satisfying $m l - nk =1$, so that $\xi$ and
$\chi$ form a basis for the periodic identifications equivalent to the
original $\partial_\phi$, $\partial_\psi$ basis. That is, in terms of
new angular coordinates $\tilde \phi, \tilde \psi$ such that $\xi =
\partial_{\tilde \phi}$, $\chi = \partial_{\tilde \psi}$, the
identifications are $\tilde \phi \sim \tilde \phi + 2\pi$, $\tilde
\psi \sim \tilde \psi + 2 \pi$.

We will need to be able to choose a gauge so that $A \cdot \xi = 0$ at
the origin. A minimal requirement is that $A \cdot \xi$ be independent
of $\theta$ at $r^2=-r_0^2$. This requires
\begin{equation}
\frac{n}{m} = \frac{a (a^2 - r_0^2) \Xi_b}{b (b^2 - r_0^2) \Xi_a}. 
\end{equation}
A smooth origin requires $\xi =0$ at $r^2=-r_0^2$, which can be decomposed
as $\xi \cdot \xi = 0$, $\xi \cdot \partial_t = 0$ and $\xi \cdot
\chi = 0$. We also require $\tilde \Delta_r(r^2=-r_0^2)=0$. These
conditions are all satisfied if 
\begin{equation} \label{c1}
(a^2-r_0^2)(b^2- r_0^2) + ab q =0
\end{equation}
and
\begin{equation} \label{c2}
q (a^2 + b^2 - r_0^2 + a^2 b^2 g^2 ) + 2ab M = 0.
\end{equation}
It is really remarkable that all these conditions are satisfied by
imposing just two constraints on the parameters: each of the three
conditions on $\xi$ involves a function of both $r$ and $\theta$.
These conditions ensure that at $r=r_0$, an $S^1$ goes to zero size.
This will be a smooth origin if
\begin{equation}
\frac{m^2 r_0^2 \delta^2}{\Xi_a^2 a^2 b^4 (b^2 - r_0^2)^2} = 1,
\end{equation}
where
\begin{equation} \label{delta}
\delta \equiv r_0^4 - 2 r_0^2 (a^2+b^2 -a^2 b^2 g^2 ) + a^4 + b^4
+a^2 b^2 -  a^2 b^2 g^2 (a^2 + b^2).
\end{equation}
To satisfy this condition, we must have $r_0^2 >0$, as stated earlier.
Returning to the gauge field, a globally well-defined gauge
transformation can be made to set $A \cdot \xi = 0$ at $r=r_0$ if
\begin{equation}
\frac{q a m}{\Xi_a (b^2- r_0^2)} = \frac{p}{g} 
\end{equation}
for some integer $p$. As before, from the ten-dimensional point of
view, $p$ specifies the twisting of the degenerating circle over the
$S^5$. The five-dimensional spacetime will have a well-defined spin
structure if $m+n$ is odd, and the ten-dimensional spacetime will if
$m+n+p$ is odd. 

These five conditions then fix the parameters $r_0,a,b,q,M$ of the
general solution in terms of three integers $m,n,p$. More explicitly,
$r_0$, $a$ and $b$ are given by solving 
\begin{equation} \label{int1}
m = \frac{ \Xi_a a b^2 (b^2 - r_0^2)}{r_0 \delta}, \quad
n= \frac{ \Xi_b a^2 b (a^2 - r_0^2)}{r_0 \delta},
\end{equation}
and
\begin{equation} \label{int2}
p = \frac{a b g (a^2 - r_0^2)(b^2 - r_0^2)}{ r_0 \delta}.
\end{equation}
$M$ and $q$ will then be given by
\begin{equation} \label{meq}
M = \frac{(a^2 - r_0^2)(
  b^2 - r_0^2)(a^2+b^2-r_0^2+a^2b^2 g^2)}{2a^2 b^2},
\end{equation}
\begin{equation} \label{qeq}
q = - \frac{(a^2-r_0^2)(b^2- r_0^2)}{ab}. 
\end{equation}

 We should check what further restrictions on the parameters
are imposed by general regularity requirements. The metric must be
regular for $r^2 > -r_0^2$, which requires $ \tilde \Delta_r
>0$, $\rho^2 > 0$, and $\Delta_\theta > 0$ for all $r^2 > -r_0^2$.
Requiring $\rho^2 >0$ implies
\begin{equation}
a^2 -r_0^2 >0, \quad  b^2 - r_0^2 >0. 
\end{equation}
Requiring $\Delta_\theta >0$ implies
\begin{equation}
\Xi_a >0, \quad \Xi_b >0. 
\end{equation}
Requiring that $-r_0^2$ is the largest root of $\tilde \Delta_r$
implies $\delta >0$. There is no obvious inconsistency between these
inequalities and (\ref{int1},\ref{int2}), so we imagine that for
generic values of the integers, we can choose a solution of
(\ref{int1},\ref{int2}) which satisfies these conditions. We note that
they automatically imply that $M >0$.

Requiring $\tilde \Delta_r >0$ also ensures that there are no horizons in the
spacetime, as it ensures that the surfaces of constant $r$ are
timelike for $r^2 > -r_0^2$. To show that the spacetime does not
contain closed timelike curves, we calculate
\begin{eqnarray}
g^{tt} &=& - \frac{(r^2+r_0^2)}{\tilde \Delta_r \Delta_\theta \rho^2 a^2
  b^2} \left\{ a^2 b^2 \Xi_a \Xi_b \left[ (r^2 + r_0^2)^2 + (\rho_0^2
    - 2r_0^2 + a^2 + b^2) (r^2 + r_0^2) \right] \right. \\ && +
\left. \delta [ \Xi_b a^2
  (b^2- r_0^2) \sin^2 \theta + \Xi_a b^2 (a^2- r_0^2) \cos^2 \theta]
\right \}, \nonumber
\end{eqnarray}
where we have used (\ref{meq},\ref{qeq}) in simplifying the form of
$g^{tt}$, and $\rho_0^2 \equiv \rho^2(r^2 = -r_0^2)$. Thus, 
$g^{tt} < 0$ for all $r^2 \geq -r_0^2$ (recall that there is an $r^2+
r_0^2$ factor in $\tilde \Delta_r$, so $g^{tt}$ is strictly less than
zero at $r^2 = -r_0^2$). That is, $dt$ is an everywhere timelike
one-form, and $t$ is a global time function. These solitons
are stably causal and hence free of closed timelike curves. Again, we
have not attempted to check for ergoregions, which would be a tedious
exercise in this case.

Finally, we consider the geometry of the surface at $r^2=-r_0^2$, with
coordinates $t, \theta, \tilde \psi$, where $\tilde \psi$ is the
$2\pi$ periodic coordinate associated with $\chi = k \partial_\phi + l
\partial_\psi$. At $r^2 = -r_0^2$,
\begin{equation}
\chi \cdot \chi = \frac{r_0^2 \delta^2 \sin^2 \theta \cos^2
  \theta}{a^4 b^4 \rho_0^4 \Xi_a^2 \Xi_b^2} [ a^2 (b^2- r_0^2) \Xi_b
\sin^2 \theta + b^2 (a^2- r_0^2) \Xi_a \cos^2 \theta],
\end{equation}
where we have used (\ref{meq},\ref{qeq}) and $lm-kn =1$. This circle
degenerates at $\theta=0$ and at $\theta = \pi/2$. At $\theta = 0$,
$\rho_0^2 = a^2 - r_0^2$ and $\Delta_\theta = \Xi_a$, so 
\begin{equation}
ds^2 \approx \ldots + \frac{a^2- r_0^2}{\Xi_a} \left(d\theta^2 + \sin^2
\theta \frac{d\tilde \psi^2}{n^2} \right). 
\end{equation}
Similarly, at $\theta = \pi/2$,
$\rho_0^2 = b^2 - r_0^2$ and $\Delta_\theta = \Xi_b$, so 
\begin{equation}
ds^2 \approx \ldots + \frac{b^2- r_0^2}{\Xi_b} \left(d\theta^2 + \sin^2
\theta \frac{d\tilde \psi^2}{m^2} \right). 
\end{equation}
The geometry will therefore have a $\mathbb{Z}_{|n|}$ orbifold
singularity at $r^2 = -r_0^2, \theta =0$ and a $\mathbb{Z}_{|m|}$
orbifold singularity at $r^2 = -r_0^2, \theta = \pi/2$.\footnote{These
  orbifold singularities appear only in the surface at $r^2=-r_0^2$.
  At other values of $r$, the $\phi$ circle shrinks as $\theta \to 0$
  and $\psi$ circle shrinks as $\theta \to \pi/2$, and these are both
  smooth origins.} That is, the only truly smooth case is $m^2=n^2=1$.

Let us consider in particular the supersymmetric case. The BPS condition for
these solutions is~\cite{Chong:2005hr}
\begin{equation}
\frac{M}{q} = 1 + (a+b) g. 
\end{equation}
Using (\ref{meq},\ref{qeq}), this can be rewritten as
\begin{equation}
r_0^2 = (a+b+ab g)^2,
\end{equation}
and \eqref{meq} becomes
\begin{equation}
M = -(1+ag+bg)(1+a g)(1+bg)(2a+b+abg)(2b+a+abg),
\end{equation}
as in~\cite{Chong:2005hr}. The integer conditions
(\ref{int1},\ref{int2}) are then
\begin{equation} \label{umc}
m = -\frac{(1-ag)(2b+a+abg)}{(a+b+abg)(3+5ag+5bg+3abg^2)},
\end{equation}
\begin{equation} \label{unc}
n = -
\frac{(1-bg)(2a+b+abg)}{(a+b+abg)(3+5ag+5bg+3abg^2)}, 
\end{equation}
and
\begin{equation} \label{upc}
p = -\frac{g (2b+a+abg)(2a+b+abg)}{(a+b+abg)(3+5ag+5bg+3abg^2)}.
\end{equation}
There are thus three conditions on the two unknowns $a,b$.  Although
we have not made a systematic search, it seems unlikely that there
will be any supersymmetric solitons. In the special case $m^2=n^2=1$,
when the geometry is completely smooth, the first two equations reduce
to $a^2=b^2$, so the analysis of section~\ref{equal} shows that there
are no supersymmetric solitons with $m^2=n^2=1$.
In~\cite{Chong:2005hr}, \eqref{umc} was imposed with $m=-1$, but the
other conditions were not. To obtain a smooth metric, we must also
impose \eqref{unc} for some value of $n$, and to make the gauge
potential well-defined we must also satisfy \eqref{upc}.

\section{Asymptotically Poincar\'e solutions}
\label{poinc}

So far, we have considered solitons which asymptotically approach AdS
in global coordinates. However, it is also possible to consider
solitons which asymptotically approach AdS in Poincar\'e coordinates
with some direction compactified, as in the AdS soliton
of~\cite{Horowitz:1998ha}. We will see that a large mass limit of our
global AdS solitons gives analogues of the AdS soliton with non-zero
gauge fields. We can exploit the twisting on the sphere to allow
periodic boundary conditions for the fermions on the compact circle in
Poincar\'e coordinates. However, we again fail to find any
supersymmetric solitons.

We analyse this for the case of unequal charges and equal angular
momenta.\footnote{We do not see any easy way to extend this scaling
  argument to the metric with equal charges and unequal angular
  momenta considered in the previous section. For an asymptotically
  Poincar\'e solution, we only expect to need one parameter for the
  momentum along the flat directions, so it may be that the solution
  obtained here is sufficiently general.} Consider the solutions
(\ref{uncmet},\ref{uncgauge}) and make the scaling\footnote{Note that
  this scaling is different from the one used in
  e.g.~\cite{Witten:1998zw,cejm} to obtain toroidal black holes. This
  scaling is dictated by the regularity conditions
  (\ref{bcond},\ref{percond}).}
\begin{equation}
\ell_1 = \gamma \bar \ell_1, \quad \ell_2 = \gamma \bar \ell_2, \quad q_i
\mbox{ fixed}, 
\end{equation}
and scale the coordinates as
\begin{equation}
  \rho = \gamma \bar \rho, \quad t = \frac{\bar t}{\gamma}, \quad \theta =
  \frac{2x}{\gamma}, \quad \psi+\phi = 
  \frac{2z}{\gamma}.
\end{equation}
Then when we take $\gamma \to \infty$, we obtain a solution
\begin{equation} \label{tormet}
\mathrm{d}s^2= -R g^2 \mathrm{d}\bar t^2 +
\frac{R\bar \rho^2}{g^2
  f_1}\mathrm{d}\bar \rho^2 +
R(dx^2 + x^2 d\phi^2) + \frac{f_{1}}{R^2} dz^2,
\end{equation}
with
\begin{eqnarray}
f_1 &=& R^3 + \frac{\bar \ell_2^6}{\bar \ell_1^2} \bar \rho^2 - \bar \ell_2^6, 
\end{eqnarray}
and $R^3 = \bar{H}_1 \bar{H}_2 \bar{H}_3$, where
\begin{equation}
\bar{H}_i = \bar \rho^2 + \bar \ell_1^2 q_i.
\end{equation}
The gauge field becomes
\begin{equation} 
  A^i = \sqrt{q_i+1} \frac{\bar \ell_2^3}{(\bar \rho^2 +
    \bar \ell_1^2 q_i)} dz, 
\end{equation}
and the scalars are 
\begin{equation}
e^{\frac{2}{\sqrt{6}} \varphi_1} = X_3, \quad e^{\sqrt{2} \varphi_2} =
  \frac{X_2}{X_1}, \quad X_i = \frac{R}{\bar{H}_i}. 
\end{equation}
Note that the form of the metric simplifies because $Y \approx g^2
f_1$ in the limit and the $f_2dt/f_1 $ term becomes negligible. This
family of geometries can also be obtained by a double analytic
continuation of charged toroidal black
holes~\cite{Behrndt:1998jd,Cvetic:1999ne}.

We would like to see what solitons can be obtained from this family.
The limiting procedure gave us the solution~\eqref{tormet} with $z \in
(-\infty,\infty)$, but to obtain smooth solitons we must make periodic
identifications $z \sim z + \Delta z$. This solution appears to depend
on five parameters, $\bar \ell_1$, $\bar \ell_2$, and $q_i$. However,
two of these parameters are coordinate degrees of freedom. This can be
made manifest by making the coordinate transformation
\begin{equation}
  r^2 = \frac{\bar \ell_1}{\bar \ell_2^3} (\bar \rho^2 -\ell_1^2),
  \quad \tau = \frac{\bar t \ell_2^{3/2}}{\ell_1^{1/2}}, \quad X = \frac{x \ell_2^{3/2}}{\ell_1^{1/2}}, \quad Z =
  \frac{z \ell_2^{3/2}}{\ell_1^{1/2}}
\end{equation}
so that 
\begin{equation} 
\mathrm{d}s^2= -R g^2 \mathrm{d}\tau^2 +
\frac{Rr^2}{g^2
  f_1}\mathrm{d}r^2 +
R(dX^2 + X^2 d\phi^2) + \frac{f_{1}}{R^2} dZ^2,
\end{equation}
with 
\begin{eqnarray}
f_1 &=& R^3 +  r^2
\end{eqnarray}
and 
\begin{equation}
\bar{H}_i = r^2 + \frac{\bar \ell_1^{3}}{\bar \ell_2^3} (q_i+1).
\end{equation}
The metric in this coordinate system depends only on the combinations
$\frac{\bar \ell_1^{3}}{\bar \ell_2^3} (q_i+1)$, so there are really
only three independent parameters. Rather than work in this new
coordinate system, it is more convenient to fix this gauge freedom by
choosing $\bar \ell_2^6 = \bar \ell_1^6 q_1 q_2 q_3$, and fixing the
period of $\Delta z$.  We reiterate that for these asymptotically
Poincar\'e solutions, this is a choice of gauge, not a regularity
condition.

Let us now consider the regularity conditions, working with the radial
coordinate $\bar \rho^2$. To have a regular origin at $\bar \rho^2 =
\bar \rho_0^2$ in the geometry~\eqref{tormet} only requires that
$f_1(\bar \rho_0^2) = 0$ and that the circle have the correct proper
radius. With the choice $\bar \ell_2^6 = \bar \ell_1^6 q_1 q_2 q_3$,
\begin{equation}
f_1 = \bar \rho^6 + \bar \rho^4 \bar \ell_1^2 (q_1 + q_2 + q_3) + \bar
\rho^2 \bar \ell_1^4 (q_1 q_2 + q_1 q_3 + q_2 q_3 + q_1 q_2 q_3), 
\end{equation}
so there is a root of $f_1$ at $\bar \rho^2 =0$. If $\bar \ell_1^2 q_i >0$,
$\bar{H}_i(\bar \rho^2 =0) >0$, so there is no singularity at larger
$\bar \rho$, and this will be the largest root of $f_1$. We use the
coordinate freedom to set $\Delta z = 2\pi$ to make this look as much
as possible like the global AdS case: the condition to have a regular
origin is then
\begin{equation}
  \frac{(q_1 q_2 + q_1 q_3 + q_2 q_3 + q_1 q_2 q_3)^2 g^2 \bar
    \ell_1^2}{q_1 q_2 q_3}  = 1.  
\end{equation}
The regularity conditions for the gauge fields are
\begin{equation}
  \frac{q_j q_k (1+q_i)}{q_i} = \frac{m_i^2}{g^2 \bar \ell_1^2}. 
\end{equation}
We see that we obtain a very similar family of smooth solitons to the
ones we had in the global AdS case. There is a discrete family of
geometries labelled by the integers $m_i$, none of which are
supersymmetric.

These asymptotically Poincar\'e solitons will arise as the near-core
limit of asymptotically flat solitons carrying D3-brane charge, just
as global AdS$_3 \times S^3$ arises as the near-core limit of the
asymptotically flat six-dimensional solitons
in~\cite{Balasubramanian:2000rt,Maldacena:2000dr}. The relevant
asymptotically flat solutions can be constructed by double analytic
continuation of rotating black D3-brane solutions. 

From the ten-dimensional perspective, the circle which shrinks at
$\bar \rho^2=0$ is $z$ twisted $m_i$ times over the three $U(1)$
angular directions in the $S^5$. Thus, if $m_1 + m_2 + m_3$ is odd,
the ten-dimensional soliton has a spin structure where the fermions
are periodic around the $z$ direction at fixed position on the sphere.
The absence of a supersymmetric soliton in this case may somehow be
connected with the need to pick out one of the flat directions to play
the role of the degenerating circle.

\section{Asymptotically flat limit}
\label{gto0}

We have found non-supersymmetric asymptotically AdS solitons in
five-dimensional gauged supergravity. We would like to understand the
relation to the asymptotically flat solitons in~\cite{nonbps}: the
three-charge metrics in that paper can be Kaluza-Klein reduced to give
asymptotically flat solutions in five-dimensional ungauged
supergravity, and one might expect that they would correspond to a $g
\to 0$ limit of the global AdS solutions.  Comparing the qualitative
properties of the two families of solitons, we see that there are some
important differences between the two cases: in the present paper, we
have found solitons for equal or unequal angular momenta, while the
asymptotically flat solitons always have unequal angular momenta. The
asymptotically AdS solitons are labelled by integers which specify all
the parameters including the charges, whereas in the asymptotically
flat case, we were free to choose the charges, and for each choice of
charges, there was a family of solitons labelled by integers. The
asymptotically AdS solitons do not have ergoregions, while the
asymptotically flat solitons do.

Despite these differences, we can make a close contact between the
limit as $g \to 0$ of the analysis in section~\ref{uneqang} and the
analysis in~\cite{nonbps}.  However, the asymptotically flat solitons
will be a distinct family which only exist when $g=0$.

We cannot take such a limit for the solitons with equal angular
momentum, as the condition $g^2 \ell_1^2 > 1$ would require us to take
some of the parameters to infinity. There is no similar condition in
the case with unequal angular momenta, however, and if we take the
limit $g \to 0$ for the metric~\eqref{uneqmet}, we obtain an
asymptotically flat solution, which for ease of comparison
to~\cite{nonbps} we rewrite as
\begin{eqnarray}
ds^2 &=& - \frac{F}{\rho^4} \left[ dt + \frac{(2M a + qb) \rho^2 - q^2
    a}{F} \sin^2 \theta d\phi + \frac{(2M b + qa) \rho^2 - q^2 b}{F}
  \cos^2 \theta d\psi \right]^2 \nonumber \\
&&+ \rho^2 \left\{ \frac{r^2 dr^2}{\tilde \Delta_r} + d\theta^2 + \sin^2 \theta
  d\phi^2 + \cos^2 \theta d\psi^2 \right. \nonumber \\
&&  + \frac{1}{F} \left[ ((a^2-b^2) \rho^2 + (2M b^2 + 2q ab)
    )\sin^4 \theta d\phi^2 \right. \nonumber \\
&& + (2Mab + q(a^2+b^2)) \sin^2 \theta \cos^2
    \theta d\phi d\psi \nonumber \\
&& \left. \left. +  ((b^2-a^2) \rho^2 + (2M a^2 + 2q ab)
    )\cos^4 \theta d\psi^2 \right] \phantom{\frac{1}{2}} \right\}, 
\end{eqnarray}
where $\rho^2$ is given in \eqref{rhosq}, $F \equiv \rho^4 - 2M \rho^2 +
q^2$, and once we have set $g = 0$, 
\begin{equation}
\tilde \Delta_r = (r^2+a^2)(r^2+b^2) + q^2 + 2abq - 2Mr^2.
\end{equation}
This is the same as the five-dimensional form of the general solution
considered in~\cite{nonbps} (for all three charges equal). To bring it
to the form used in~\cite{nonbps}, we should redefine the parameters
by
\begin{equation}
2M = \bar M (\cosh^2 \delta + \sinh^2 \delta), \quad q = -\bar M \sinh
\delta \cosh \delta ,
\end{equation}
\begin{equation}
a = a_1 \sinh \delta - a_2 \cosh \delta, \quad b = a_1 \cosh \delta -
a_2 \sinh \delta, 
\end{equation}
and define a new radial coordinate 
\begin{equation}
\bar r^2 = r^2 + (a_1^2+a_2^2) \sinh^2 \delta - 2 a_1 a_2 \sinh
\delta \cosh \delta - \bar M \sinh^2 \delta. 
\end{equation}

To understand the relation between the asymptotically AdS and
asymptotically flat solitons, we would like to re-express the
conditions (\ref{c1},\ref{c2}) and the integers
(\ref{int1},\ref{int2}) in terms of these new parameters. We can think
of \eqref{c2} as determining $r_0^2$; in terms of the new parameters,
it gives
\begin{equation}
r_0^2 = \frac{a_1 a_2}{\sinh \delta \cosh \delta}. 
\end{equation}
Substituting this into \eqref{c1} gives
\begin{equation}
\bar M = a_1^2 + a_2^2 - a_1 a_2 \frac{\cosh^6 \delta + \sinh^6
  \delta}{\cosh^3 \delta \sinh^3 \delta},
\end{equation}
which matches (3.15) in~\cite{nonbps}. The expressions for the
integers $m$ and $n$ in~\eqref{int1} become
\begin{equation}
m = \frac{(\sinh \delta \cosh \delta)^{3/2} (a_1 \cosh^3 \delta - a_2
\sinh^3\delta)}{\sqrt{a_1 a_2} (\cosh^6 \delta - \sinh^6 \delta)},
\end{equation}
\begin{equation}
  n =  \frac{(\sinh \delta \cosh \delta)^{3/2} (a_1
    \sinh^3\delta- a_2 \cosh^3 \delta)}{\sqrt{a_1 a_2} (\cosh^6 \delta
    - \sinh^6 \delta)}. 
\end{equation}
These are thus the same as the integers $m$ and $n$ labelling the
solitons in \cite{nonbps}. When we set $g=0$ in \eqref{int2}, on the
other hand, it simply reduces to
\begin{equation}
p =0. 
\end{equation}
This is an allowed value for $p$, from the point of view of the
integer quantisation. Thus, we recover the asymptotically flat
solitons of~\cite{nonbps} as the special case $g=0$, $p=0$ of the
solutions in the previous section.\footnote{As noted above, the
  five-dimensional geometry has orbifold singularities at $r^2 =
  -r_0^2$, $\theta = 0, \pi/2$ for general values of $m,n$. The
  asymptotically flat geometry is completely smooth only in
  six dimensions.} Because \eqref{int2} is automatically
satisfied, we have one free parameter, which we can use to choose the
charge arbitrarily.

Note that $p \propto mn g$, so for $g \neq 0$ we cannot set $p=0$, as
it would contradict our assumption that $mn \neq 0$. Thus, the
asymptotically flat solitons are a distinct family which appear when
$g = 0$, and not the special case $g=0$ of the asymptotically AdS
solitons.  This explains why the properties of the asymptotically AdS
and asymptotically flat cases are different.

In addition, while there is a close relation between the two families
as solutions in five dimensions, their interpretation in string theory
is quite different: the gauged supergravity solutions arise from type
IIB compactified on $S^5$, with a nonzero five-form field strength
proportional to $g$, and angular momenta on the $S^5$ proportional to
$q$. The asymptotically flat solutions arise for example from type IIB
compactified on $T^5$, with a three-form field strength proportional
to $q$ and a momentum proportional to $q$ on one of the $T^5$
directions.

\section{Conclusions}
\label{concl}

In this paper, we have found non-supersymmetric solitons in
five-dimensional gauged supergravity. Surprisingly, we were unable to
find any supersymmetric solitons.  The solutions have a circle which
degenerates smoothly in the interior of the spacetime. As in previous
studies of asymptotically flat and AdS$_3 \times S^3$ cases, a
twisting plays a crucial role: the circle which degenerates in the
interior is twisted over the $S^5$ in the ten-dimensional metric.
These non-supersymmetric solitons have a different character to the
1/2 BPS asymptotically AdS$_5 \times S^5$ geometries found
in~\cite{llm}: In particular, they have a non-contractible $S^2$, so
they have a different topology.

There are a number of directions in which this work could be extended.
These solitons will have corresponding states in the dual
$\mathcal{N}=4$ super Yang-Mills
theory~\cite{Maldacena:1997re,Aharony:1999ti}. One can read off the
charges of these states and the expectation values of field theory
operators from the bulk solutions given above, but explicitly
identifying the corresponding field theory states in the absence of
supersymmetry will be challenging, and we currently have no clear idea
how to proceed.\footnote{It may be possible to relate the twisting of
  the degenerating circle over the $S^5$ to something like spectral
  flow in the dual field theory, but this is probably not the whole
  story.}  Nonetheless, this is an important problem to address: it is
hard to imagine we will get much simpler examples of
non-supersymmetric solitons to work with, and extending the AdS/CFT
dictionary beyond the supersymmetric context is important for a host
of reasons.

It would also be interesting to compare the properties of these
soliton geometries to those of the charged rotating black holes, to
explore if the solitons could be interpreted as microstates of the
black holes following the ideas of~\cite{mrev}. Another sense in which
solitons and black holes could be compared is,
following~\cite{Horowitz:2005vp,Ross:2005ms}, to ask if there are
black holes with a winding tachyon whose condensation could
lead to one of these solitons.

Their stability should also be investigated.  We showed in
section~\ref{equal} that the solitons with equal charges and equal
angular momenta had no ergoregions. We did not explore the existence
of ergoregions for the other cases; it was enough to observe that we
had some examples in the asymptotically AdS case of non-supersymmetric
soliton solutions without ergoregions (unlike in the asymptotically flat
case).  Hence, these solutions can avoid the instability argument
of~\cite{fergo}. This of course does not imply that they are stable;
further study is necessary.

We will finally remark on the extension of this analysis to different
numbers of dimensions. Similar families of charged rotating black hole
solutions have been found for the gauged supergravity theories in four
and seven
dimensions~\cite{Carter:1968ks,Carter:1968bb,Chong:2004na,Chong:2004dy},
and we would like to find solitons in these
theories as well.  It is clear that in the four-dimensional case, it
will not be possible to construct smooth solitons in the same way.
There is only one $U(1)$ direction in the transverse two-sphere in the
four-dimensional solution, and it shrinks to zero size at the north
and south poles of the two-sphere. Thus, its size is always a function
of $\theta$, and it will not be possible to combine it with the radial
coordinate to form a smooth origin for all $\theta$.

The seven-dimensional case seems at first blush more promising. The
spacetime involves an $S^5$, and the metric is naturally written in
coordinates that write the $S^5$ as an $S^1$ bundle over
$\mathbb{CP}^2$~\cite{Chong:2004dy},
\begin{equation}
ds^2 = \ldots + A( d\xi^2 + \frac{1}{4} \sin^2 \xi (\sigma_1^2 + \sigma_2^2)
+ \frac{1}{4} \sin^2 \xi \cos^2 \xi \sigma_3^2)  + B (\sigma + C dt)^2, 
\end{equation}
where the $\sigma_i$ are the left-invariant one-forms on $S^3$
(\ref{sigma1}-\ref{sigma3}), and
\begin{equation}
\sigma = d\tau + \frac{1}{2} \sin^2 \xi \sigma_3.
\end{equation}
It therefore seems natural to look for smooth solitons where
the radial direction combines with the $\tau$ circle to form a smooth
origin. Supersymmetric solutions with smooth metrics were indeed found
in~\cite{Cvetic:2005zi}. However, there is a problem with the gauge
fields. In the charged black holes of~\cite{Chong:2004dy},
the solutions involve a non-trivial three-form potential
\begin{equation}
A_{(3)} = \frac{2 m a s^2}{\Xi \Xi_- (r^2 + a^2)} \sigma \wedge J,
\end{equation}
where $J$ is the K\"ahler form on $\mathbb{CP}^2$. If we consider the
holonomy of this three-form obtained by integrating it over the $\tau$
circle and an $S^2 \subset \mathbb{CP}^2$ parametrised by
$\theta,\phi$,
\begin{equation}
\oint_{S^1 \times S^2} A_{(3)} = \frac{2 m a s^2}{\Xi \Xi_- (r^2 +
  a^2)} 2\pi^2 \sin^2 \xi. 
\end{equation}
Since this holonomy depends on the coordinate $\xi$ labelling the
two-sphere we choose, it is impossible to choose parameters such that
this holonomy is trivial for all the possible cycles. (Setting $ma s^2
= 0$ would set either the charge or the angular momentum to zero, and
there are no solutions with smooth metrics for these cases.) Thus, we
cannot find any solitons with smooth three-form in this
family of solutions.  

\medskip
{\large\bf Acknowledgements}

I thank Owen Madden for collaboration on early stages of this
project. I am also grateful for useful discussions with Mirjam Cveti\v{c}, Neil
Lambert and Don Marolf. This work is supported by the EPSRC. 

\bibliographystyle{/home/aplm/dma0sfr/tex_stuff/bibs/utphys}

\bibliography{unique}

\providecommand{\href}[2]{#2}\begingroup\raggedright\begin{thebibliography}{10}

\bibitem{Balasubramanian:2000rt}
V.~Balasubramanian, J.~de~Boer, E.~Keski-Vakkuri, and S.~F. Ross,
  ``Supersymmetric conical defects: Towards a string theoretic description of
  black hole formation,'' Phys. Rev. D {\bf 64} (2001) 064011,
\href{http://xxx.lanl.gov/abs/hep-th/0011217}{{\tt hep-th/0011217}}.

\bibitem{Maldacena:2000dr}
J.~M. Maldacena and L.~Maoz, ``De-singularization by rotation,'' JHEP {\bf 12}
  (2002) 055,
\href{http://xxx.lanl.gov/abs/hep-th/0012025}{{\tt hep-th/0012025}}.

\bibitem{llm}
H.~Lin, O.~Lunin, and J.~Maldacena, ``Bubbling {AdS} space and 1/2 {BPS}
  geometries,'' JHEP {\bf 10} (2004) 025,
\href{http://xxx.lanl.gov/abs/hep-th/0409174}{{\tt hep-th/0409174}}.

\bibitem{Cvetic:2005zi}
M.~Cveti\v{c}, G.~W. Gibbons, H.~L\"u, and C.~N. Pope, ``Rotating black holes
  in gauged supergravities: Thermodynamics, supersymmetric limits, topological
  solitons and time machines,''
\href{http://xxx.lanl.gov/abs/hep-th/0504080}{{\tt hep-th/0504080}}.

\bibitem{Chong:2005da}
Z.~W. Chong, M.~Cveti\v{c}, H.~L\"u, and C.~N. Pope, ``Five-dimensional gauged
  supergravity black holes with independent rotation parameters,'' Phys. Rev. D
  {\bf 72} (2005) 041901,
\href{http://xxx.lanl.gov/abs/hep-th/0505112}{{\tt hep-th/0505112}}.

\bibitem{Chong:2005hr}
Z.~W. Chong, M.~Cveti\v{c}, H.~L\"u, and C.~N. Pope, ``General non-extremal
  rotating black holes in minimal five- dimensional gauged supergravity,''
\href{http://xxx.lanl.gov/abs/hep-th/0506029}{{\tt hep-th/0506029}}.

\bibitem{nonbps}
V.~Jejjala, O.~Madden, S.~F. Ross, and G.~Titchener, ``Non-supersymmetric
  smooth geometries and {D1-D5-P} bound states,'' Phys. Rev. D {\bf 71} (2005)
  124030,
\href{http://xxx.lanl.gov/abs/hep-th/0504181}{{\tt hep-th/0504181}}.

\bibitem{cy}
M.~Cveti\v{c} and D.~Youm, ``General rotating five dimensional black holes of
  toroidally compactified heterotic string,'' Nucl. Phys. {\bf B476} (1996)
  118--132,
\href{http://xxx.lanl.gov/abs/hep-th/9603100}{{\tt hep-th/9603100}}.

\bibitem{lm1}
O.~Lunin and S.~D. Mathur, ``Metric of the multiply wound rotating string,''
  Nucl. Phys. {\bf B610} (2001) 49--76,
\href{http://xxx.lanl.gov/abs/hep-th/0105136}{{\tt hep-th/0105136}}.

\bibitem{lm2}
O.~Lunin and S.~D. Mathur, ``{AdS/CFT} duality and the black hole information
  paradox,'' Nucl. Phys. {\bf B623} (2002) 342--394,
\href{http://xxx.lanl.gov/abs/hep-th/0109154}{{\tt hep-th/0109154}}.

\bibitem{lmm}
O.~Lunin, J.~Maldacena, and L.~Maoz, ``Gravity solutions for the {D1-D5} system
  with angular momentum,''
\href{http://xxx.lanl.gov/abs/hep-th/0212210}{{\tt hep-th/0212210}}.

\bibitem{Lunin:2004uu}
O.~Lunin, ``Adding momentum to {D1-D5} system,'' JHEP {\bf 04} (2004) 054,
\href{http://xxx.lanl.gov/abs/hep-th/0404006}{{\tt hep-th/0404006}}.

\bibitem{Giusto:2004id}
S.~Giusto, S.~D. Mathur, and A.~Saxena, ``Dual geometries for a set of 3-charge
  microstates,'' Nucl. Phys. {\bf B701} (2004) 357--379,
\href{http://xxx.lanl.gov/abs/hep-th/0405017}{{\tt hep-th/0405017}}.

\bibitem{Giusto:2004ip}
S.~Giusto, S.~D. Mathur, and A.~Saxena, ``3-charge geometries and their {CFT}
  duals,'' Nucl. Phys. {\bf B710} (2005) 425--463,
\href{http://xxx.lanl.gov/abs/hep-th/0406103}{{\tt hep-th/0406103}}.

\bibitem{Bena:2005va}
I.~Bena and N.~P. Warner, ``Bubbling supertubes and foaming black holes,''
\href{http://xxx.lanl.gov/abs/hep-th/0505166}{{\tt hep-th/0505166}}.

\bibitem{Berglund:2005vb}
P.~Berglund, E.~G. Gimon, and T.~S. Levi, ``Supergravity microstates for {BPS}
  black holes and black rings,''
\href{http://xxx.lanl.gov/abs/hep-th/0505167}{{\tt hep-th/0505167}}.

\bibitem{clp1}
M.~Cveti\v{c}, H.~L\"u, and C.~N. Pope, ``Charged {Kerr-de S}itter black holes
  in five dimensions,'' Phys. Lett. {\bf B598} (2004) 273--278,
\href{http://xxx.lanl.gov/abs/hep-th/0406196}{{\tt hep-th/0406196}}.

\bibitem{clp2}
M.~Cveti\v{c}, H.~L\"u, and C.~N. Pope, ``Charged rotating black holes in five
  dimensional {$U(1)^3$ gauged $N = 2$} supergravity,'' Phys. Rev. D {\bf 70}
  (2004) 081502,
\href{http://xxx.lanl.gov/abs/hep-th/0407058}{{\tt hep-th/0407058}}.

\bibitem{Horowitz:1998ha}
G.~T. Horowitz and R.~C. Myers, ``The {AdS/CFT} correspondence and a new
  positive energy conjecture for general relativity,'' Phys. Rev. D {\bf 59}
  (1999) 026005,
\href{http://xxx.lanl.gov/abs/hep-th/9808079}{{\tt hep-th/9808079}}.

\bibitem{eh1}
R.~Clarkson and R.~B. Mann, ``Eguchi-hanson solitons,''
\href{http://xxx.lanl.gov/abs/hep-th/0508109}{{\tt hep-th/0508109}}.

\bibitem{eh2}
R.~Clarkson and R.~B. Mann, ``Eguchi-{H}anson solitons in odd dimensions,''
\href{http://xxx.lanl.gov/abs/hep-th/0508200}{{\tt hep-th/0508200}}.

\bibitem{mrev}
S.~D. Mathur, ``The fuzzball proposal for black holes: An elementary review,''
  Fortsch. Phys. {\bf 53} (2005) 793--827,
\href{http://xxx.lanl.gov/abs/hep-th/0502050}{{\tt hep-th/0502050}}.

\bibitem{uniq}
O.~Madden and S.~F. Ross, ``On uniqueness of charged {Kerr-AdS} black holes in
  five dimensions,'' Class. Quant. Grav. {\bf 22} (2005) 515--524,
\href{http://xxx.lanl.gov/abs/hep-th/0409188}{{\tt hep-th/0409188}}.

\bibitem{Cvetic:1999xp}
M.~Cveti\v{c} {\em et.~al.}, ``Embedding {AdS} black holes in ten and eleven
  dimensions,'' Nucl. Phys. {\bf B558} (1999) 96--126,
\href{http://xxx.lanl.gov/abs/hep-th/9903214}{{\tt hep-th/9903214}}.

\bibitem{Witten:1998zw}
E.~Witten, ``Anti-de {S}itter space, thermal phase transition, and confinement
  in gauge theories,'' Adv. Theor. Math. Phys. {\bf 2} (1998) 505--532,
\href{http://xxx.lanl.gov/abs/hep-th/9803131}{{\tt hep-th/9803131}}.

\bibitem{cejm}
A.~Chamblin, R.~Emparan, C.~V. Johnson, and R.~C. Myers, ``Charged {AdS} black
  holes and catastrophic holography,'' Phys. Rev. D {\bf 60} (1999) 064018,
\href{http://xxx.lanl.gov/abs/hep-th/9902170}{{\tt hep-th/9902170}}.

\bibitem{Behrndt:1998jd}
K.~Behrndt, M.~Cveti\v{c}, and W.~A. Sabra, ``Non-extreme black holes of five
  dimensional {$N = 2$ AdS} supergravity,'' Nucl. Phys. {\bf B553} (1999)
  317--332,
\href{http://xxx.lanl.gov/abs/hep-th/9810227}{{\tt hep-th/9810227}}.

\bibitem{Cvetic:1999ne}
M.~Cveti\v{c} and S.~S. Gubser, ``Phases of {R}-charged black holes, spinning
  branes and strongly coupled gauge theories,'' JHEP {\bf 04} (1999) 024,
\href{http://xxx.lanl.gov/abs/hep-th/9902195}{{\tt hep-th/9902195}}.

\bibitem{Maldacena:1997re}
J.~M. Maldacena, ``The large {N} limit of superconformal field theories and
  supergravity,'' Adv. Theor. Math. Phys. {\bf 2} (1998) 231--252,
\href{http://xxx.lanl.gov/abs/hep-th/9711200}{{\tt hep-th/9711200}}.

\bibitem{Aharony:1999ti}
O.~Aharony, S.~S. Gubser, J.~M. Maldacena, H.~Ooguri, and Y.~Oz, ``Large {N}
  field theories, string theory and gravity,'' Phys. Rept. {\bf 323} (2000)
  183--386,
\href{http://xxx.lanl.gov/abs/hep-th/9905111}{{\tt hep-th/9905111}}.

\bibitem{Horowitz:2005vp}
G.~T. Horowitz, ``Tachyon condensation and black strings,'' JHEP {\bf 08}
  (2005) 091,
\href{http://xxx.lanl.gov/abs/hep-th/0506166}{{\tt hep-th/0506166}}.

\bibitem{Ross:2005ms}
S.~F. Ross, ``Winding tachyons in asymptotically supersymmetric black
  strings,'' JHEP {\bf 10} (2005) 112,
\href{http://xxx.lanl.gov/abs/hep-th/0509066}{{\tt hep-th/0509066}}.

\bibitem{fergo}
J.~L. Friedman, ``Ergosphere instability,'' Commun. Math. Phys. {\bf 63} (1978)
  243.

\bibitem{Carter:1968ks}
B.~Carter, ``Hamilton-{J}acobi and {S}chrodinger separable solutions of
  {E}instein's equations,'' Commun. Math. Phys. {\bf 10} (1968)
280.

\bibitem{Carter:1968bb}
B.~Carter, ``A new family of {E}instein spaces,'' Phys. Lett. {\bf A26} (1968)
  399.

\bibitem{Chong:2004na}
Z.~W. Chong, M.~Cveti\v{c}, H.~L\"u, and C.~N. Pope, ``Charged rotating black
  holes in four-dimensional gauged and ungauged supergravities,'' Nucl. Phys.
  {\bf B717} (2005) 246--271,
\href{http://xxx.lanl.gov/abs/hep-th/0411045}{{\tt hep-th/0411045}}.

\bibitem{Chong:2004dy}
Z.~W. Chong, M.~Cveti\v{c}, H.~L\"u, and C.~N. Pope, ``Non-extremal charged
  rotating black holes in seven- dimensional gauged supergravity,''
\href{http://xxx.lanl.gov/abs/hep-th/0412094}{{\tt hep-th/0412094}}.

\end{thebibliography}\endgroup

\end{document}